\documentclass[10pt,twocolumn,showpacs,preprintnumbers,amsmath,amssymb,aps,prl,superscriptaddress,longbibliography]{revtex4-2}
\usepackage{appendix}
\usepackage{bbm}
\usepackage{mathrsfs}
\usepackage{graphicx}
\usepackage{dcolumn}
\usepackage{bm}
\usepackage{braket}
\usepackage{amsmath}
\usepackage{amsfonts}
\usepackage[dvipsnames]{xcolor}
\usepackage[colorlinks=true,linkcolor=Blue,urlcolor=BlueViolet,citecolor=BlueViolet]{hyperref}
\usepackage{natbib}
\usepackage{multirow}
\usepackage{floatrow}
\usepackage{hyperref}
\usepackage{tikz}

\begin{document}

\title{Flat Chern bands and correlated states in spiral magnet ReAg$_2$Cl$_6$}
\author{Kejie Bao}
\affiliation{State Key Laboratory of Surface Physics and Department of Physics, Fudan University, Shanghai 200433, China}
\affiliation{Shanghai Research Center for Quantum Sciences, Shanghai 201315, China}
\author{Rui Shi}
\affiliation{State Key Laboratory of Surface Physics and Department of Physics, Fudan University, Shanghai 200433, China}
\affiliation{Shanghai Research Center for Quantum Sciences, Shanghai 201315, China}
\author{Huan Wang}
\affiliation{State Key Laboratory of Surface Physics and Department of Physics, Fudan University, Shanghai 200433, China}
\affiliation{Shanghai Research Center for Quantum Sciences, Shanghai 201315, China}
\author{Jiaxuan Guo}
\affiliation{State Key Laboratory of Surface Physics and Department of Physics, Fudan University, Shanghai 200433, China}
\author{Jing Wang}
\thanks{Contact author: wjingphys@fudan.edu.cn}
\affiliation{State Key Laboratory of Surface Physics and Department of Physics, Fudan University, Shanghai 200433, China}
\affiliation{Shanghai Research Center for Quantum Sciences, Shanghai 201315, China}
\affiliation{Institute for Nanoelectronic Devices and Quantum Computing, Fudan University, Shanghai 200433, China}
\affiliation{Hefei National Laboratory, Hefei 230088, China}

\begin{abstract}
We predict the van der Waals monolayer ReAg$_2$Cl$_6$ hosts isolated flat Chern bands at the Fermi level in its $120^\circ$ antiferromagnetic ground state. Their flatness and nontrivial topology arise from the cooperative effect of coplanar spin order and strong spin-orbit coupling within Re $5d$ orbitals—a mechanism distinct from moir\'e systems. The spiral spin texture naturally enlarges the unit cell, reducing carrier densities while preserving sizable interaction scales. Many-body calculations show that fractional fillings can support fractional Chern insulator and charge-density wave states. Remarkably, the mechanism is generic to a broad family of Re-based compounds, with both spin configuration and flat band topology tunable by electrical manipulation. Our findings establish Re-based coplanar antiferromagnets as a robust, tunable, and experimentally accessible platform for flat Chern bands and correlated topological phases potentially at elevated temperatures.
\end{abstract}


\maketitle

Exotic correlated quantum matter emerges from the interplay between nontrivial band topology and strong electron interactions in two-dimensional (2D) materials. A paradigmatic example is the fractional Chern insulator (FCI), recently discovered in moir\'e systems at zero magnetic field~\cite{cai2023,zeng2023,park2023,xu2023,lu2024,spanton2018,xie2021,li2021,devakul2021,crepel2023,cano2023,yu2023,dong2023a,goldman2023,wang2024,jia2024,macdonald2024,song2024}. These fractional topological states originate from flat Chern minibands~\cite{tang2011,sun2011,neupert2011,qi2011,regnault2011,sheng2011}, where the moir\'e superlattice quenches the kinetic energy of dispersive bands and enhances the role of Coulomb interactions~\cite{barticevic2010,bistritzer2011,nuckolls2024}. The large moir\'e supercell also yields a low carrier density ($n \sim 10^{12}$ cm$^{-2}$), enabling precise tuning of band filling and correlated phases. However, moir\'e systems are intrinsically fragile: twist-angle inhomogeneity, modest interaction scales ($U \sim 10$ meV), and device complexity reduce the robustness of correlated states. Consequently, the fractional quantum anomalous Hall (QAH) effect has so far been observed only at cryogenic temperatures (below $1$~K)~\cite{park2023,xu2023,lu2024}, constraining potential applications~\cite{Nayak2008}. These limitations motivate the search for stoichiometric 2D materials—preferably monolayers—with intrinsic flat topological bands, which could provide more robust platforms for correlated quantum phases at higher energy scales.

\begin{center}
\tikzset{every picture/.style={line width=0.75pt}} 
\begin{tikzpicture}[x=0.5pt,y=0.5pt,yscale=-1,xscale=1]
\draw    (135.31,39.2) -- (524.8,39.6) ;
\draw [shift={(527.8,39.6)}, rotate = 180.07] [fill={rgb, 255:red, 0; green, 0; blue, 0 }  ][line width=0.08]  [draw opacity=0] (10.72,-5.15) -- (0,0) -- (10.72,5.15) -- cycle    ;
\draw [shift={(132.31,39.2)}, rotate = 0.07] [fill={rgb, 255:red, 0; green, 0; blue, 0 }  ][line width=0.08]  [draw opacity=0] (10.72,-5.15) -- (0,0) -- (10.72,5.15) -- cycle    ;
\draw  [fill={rgb, 255:red, 0; green, 0; blue, 0 }  ,fill opacity=1 ] (321.16,39.4) .. controls (321.16,37.25) and (322.9,35.5) .. (325.06,35.5) .. controls (327.21,35.5) and (328.96,37.25) .. (328.96,39.4) .. controls (328.96,41.55) and (327.21,43.3) .. (325.06,43.3) .. controls (322.9,43.3) and (321.16,41.55) .. (321.16,39.4) -- cycle ;
\draw (115,47) node [anchor=north west][inner sep=0.75pt]   [align=left] {moir\'e system};
\draw (282,47) node [anchor=north west][inner sep=0.75pt]   [align=left] {AFM system};
\draw (472,47) node [anchor=north west][inner sep=0.75pt]   [align=left] {FM system};
\draw (115,70.39) node [anchor=north west][inner sep=0.75pt]   [align=left] {\small{$U\sim10$~meV}};
\draw (282,70.39) node [anchor=north west][inner sep=0.75pt]   [align=left] {\small{$U\sim0.1$~eV}};
\draw (472,70.39) node [anchor=north west][inner sep=0.75pt]   [align=left] {\small{$U\sim0.2$~eV}};
\draw (115,92.39) node [anchor=north west][inner sep=0.75pt]   [align=left] { \small{$n\sim10^{12}$~cm$^{-2}$}};
\draw (282,92.39) node [anchor=north west][inner sep=0.75pt]   [align=left] {\small{$n\sim10^{13}$-$10^{14}$~cm$^{-2}$}};
\draw (472,92.39) node [anchor=north west][inner sep=0.75pt]   [align=left] {\small{$n\sim10^{15}$~cm$^{-2}$}};
\end{tikzpicture}
\end{center}

The realization of flat Chern bands requires a delicate balance among lattice hopping, spin-orbit coupling (SOC), and magnetism~\cite{mielke1991a,mielke1991b,wu2007,bergman2008,liu2014,ma2020,calugaru2022}. Previous efforts have focused on 2D ferromagnetic (FM) crystals~\cite{liu2021,regnault2022,liu2013,yamada2016,sun2022,pan2023,bhattacharya2023,neves2024,duan2024,ye2024,bao2025}, where the interaction scale is sizable ($U\sim0.2$~eV), but small primitive cells lead to high carrier densities ($n\sim10^{15}$~cm$^{-2}$), making band filling difficult to control. Moir\'e superlattices alleviate this constraint by enlarging the effective unit cell, yet the associated interaction scales remain modest. Antiferromagnetic (AFM) crystals provide an alternative route that combines the advantages of both approaches: spin ordering naturally enlarges the unit cell, lowering carrier densities to $n\sim10^{13}$–$10^{14}$~cm$^{-2}$—well within gate tunability—while retaining strong interactions with $U \sim 0.1$~eV, exceeding those in moir\'e systems by an order of magnitude.

2D AFM crystals have been largely overlooked because N\'eel AFMs possess trivial spin-degenerate bands, resembling nonmagnetic materials. Noncoplanar AFMs, while exhibiting intrinsic spin chirality and nontrivial band topology even without spin-orbit coupling (SOC)~\cite{ye1999,shindou2001,martin2008}, are difficult to stabilize experimentally. In contrast, coplanar AFMs—readily realized via geometric frustration on triangular lattices—represent a more practical route to flat Chern bands with sizable interaction scales. Although coplanar spin textures yield only trivial effective flux and vanishing scalar spin chirality without SOC, its inclusion fundamentally changes this picture. The real-space Berry phase of itinerant electrons in coplanar AFMs acquires an additional contribution~\cite{zhang2020}
\begin{equation}\label{eq1}
\gamma = \frac{1}{2}\int_{\mathcal{S}_c}\left[(\partial_\mu \tilde{A}_\nu - \partial_\nu \tilde{A}_\mu) +\bm{n}\cdot (\partial_{\mu}\bm{n}\times \partial_{\nu}\bm{n})\right]d^2\sigma^{\mu\nu}.
\end{equation}
Here the second term corresponds to the scalar spin chirality with local moments $\bm{S}(\mathbf{r})\equiv S\boldsymbol{n}(\mathbf{r})$, while the first term captures the synergistic effect of SOC and magnetic order. $\mathcal{S}_c$ is the area,  $A_\mu^a$ denotes the $SU(2)$ gauge field arising from SOC, and $\tilde{A}_\mu=n^a A_\mu^a$ is the projected $U(1)$ gauge field along spin direction $\bm{n}$. This synergistic contribution from SOC and spin order acts as an emergent magnetic field for electrons, thereby enabling the formation of Chern bands in coplanar AFMs.

In this Letter, we predict that monolayer ReAg$_2$Cl$_6$, a van der Waals crystal, hosts four consecutive, isolated, and flat Chern bands at the Fermi level in its coplanar AFM ground state, as revealed by density functional theory (DFT)~\cite{kresse1996,perdew1996,dudarev1998,krukau2006} and tight-binding calculations. The coplanar $120^\circ$ spin structure arises from geometric frustration on trianglular lattice, the strong SOC originates from Re $5d$  orbitals. Many-body calculations further indicates that partial filling of these Chern bands may stabilize FCI and charge density wave (CDW) state. These results demonstrate that 2D coplanar AFMs with SOC constitute a realistic and promising platform for flat Chern bands and correlated quantum phases.

\begin{table}[b]
\caption{Lattice constant; nearest-neighbor AFM exchange parameter $J$; magnetocrystalline anisotropy energy (MAE) per unit cell $E_{\text{MAE}}$, defined as the total energy difference between in-plane and out-of-plane spin configurations; N\'eel temperature $T_{\text{N}}$ from Monte Carlo simulation; band gap $E_{g}$.}
\begin{center}\label{tab1}
\renewcommand{\arraystretch}{1.4}
\begin{tabular*}{\columnwidth}
{@{\extracolsep{\fill}}cccccc}
\hline
\hline
 Material& $a$ (\AA) & $J$ (meV) &  $E_{\rm{MAE}}$ (meV) & $T_{\text{N}}$ (K) & $E_{g}$ (eV)\\
 \hline
ReAg$_2$Cl$_6$ & 6.78 & 2.35 & 0.80 & 22 & 1.24\\
\hline
\hline
\end{tabular*}
\end{center}
\end{table}

\begin{figure}[t]
\begin{center}
\includegraphics[width=\columnwidth, clip=true]{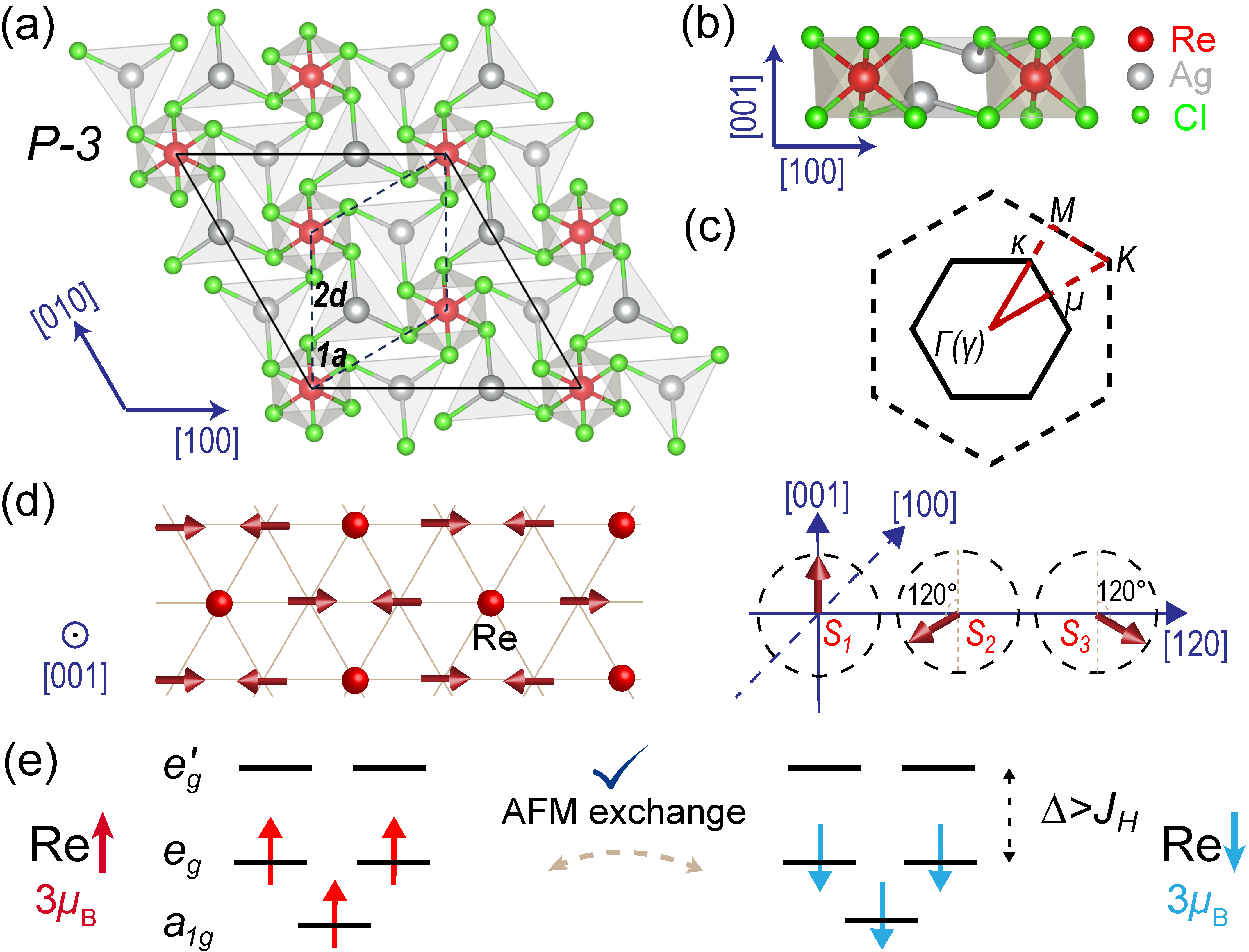}
\end{center}
\caption{(a,b) Atomic structure of monolayer ReAg$_2$Cl$_6$ from top and side views. The Wyckoff positions 1$a$ and 2$d$ are displayed (notation adopted from Bilbao Crystallographic Server~\cite{bilbao2,bilbao3,elcoro2017}). The original primitive cell and the magnetic $\sqrt{3}\times\sqrt{3}\times1$ supercell are represented as dashed and solid lines, respectively. (c) Brillouin zone (BZ) of the primitive cell and the supercell. (d) Schematic illustration of (100) AFM with 120$^{\circ}$ spin spiral structure. (e) Crystal field splitting and schematic diagram of AFM exchange between Re 5$d$ electrons.}
\label{fig1}
\end{figure}

\emph{Structure and magnetic properties---}The monolayer ReAg$_2$Cl$_6$ has a triangular lattice with the space group $P$-$3$ (No.~147). As shown in Fig.~\ref{fig1}(a), each Re atom is octahedrally coordinated with six surrounding nearest Cl anions, while Ag atom are surrounded by three Cl atoms forming [AgCl$_3$]$^{2-}$ unit, making a sandwich arrangement of Re atoms. The lattice constant is listed in Table~\ref{tab1}. The dynamical and thermal stability are confirmed by first-principles phonon and molecular dynamics calculations, respectively~\cite{supple}.  Remarkably, the van der Waals bulk ReAg$_2$Cl$_6$ has been successfully synthesized in experiments, and our calculated structure perfectly matches the X-ray diffraction result~\cite{martinez2006two}. Meanwhile, its exfoliation energy comparable to that of layered transition-metal dichalcogenides implies that monolayers can be readily exfoliated from bulk crystals~\cite{supple}.

First-principles calculations reveal that  ReAg$_2$Cl$_6$ stabilizes in a 120$^{\circ}$ spiral AFM ground state~\cite{supple,xiang2013,liuspin2024}, driven by strong nearest-neighbor AFM coupling between Re atoms in the triangular lattice. As shown in Fig.~\ref{fig1}(d), the spin-spiral lies in the $(100)$ plane (hereafter denoted as ``$(100)$ AFM''), with magnetic modulation vector $\mathbf{q}=(1/3,1/3,0)$ and a finite out-of-plane component on each moment. This magnetic ordering lowers the crystal symmetry  from $P$-$3$ to $P1$. The underlying mechanism of AFM can be elucidated from orbital occupation. Each Re atom carries a magnetic moment of about $2.9\mu_B$. Under trigonal crystal fields, the Re $5d$ orbitals split into $a_{1g} (d_{z^2})$, $e_{g}$($d_{x^2-y^2}, d_{xy}$), and $e'_{g}$($d_{xz}, d_{yz}$) [Fig.~\ref{fig1}(e)], following Mulliken notation~\cite{Georgescu2022Trigonal}. The $a_{1g}$ and $e_{g}$ states lie lower in energy than $e'_{g}$ because the latter point toward negatively charged ligands. Consequently, each Re$^{4+}$ cation adopts the $a_{1g}^{1}e_{g}^{2}e'^{0}_{g}$ configuration with the moment of $3\mu_B$ according to Hund's rule, which is close to our DFT results. Since the crystal field splitting $\Delta$ exceeds the Hund's coupling $J_{\text{H}}$ in this $5d$ system, a strong AFM exchange interaction is anticipated between neighboring Re sites~\cite{khomskii2004}, in agreement with the parameters summarized in Table~\ref{tab1}. The calculated N\'eel temperature of $\sim22$~K for the monolayer is slightly reduced compared to the bulk value in experiments ($26$~K)~\cite{martinez2006two}. Finally, the band gap listed in Table~\ref{tab1} suggests its semiconducting nature.

\begin{figure}[t]
\begin{center}
\includegraphics[width=\columnwidth, clip=true]{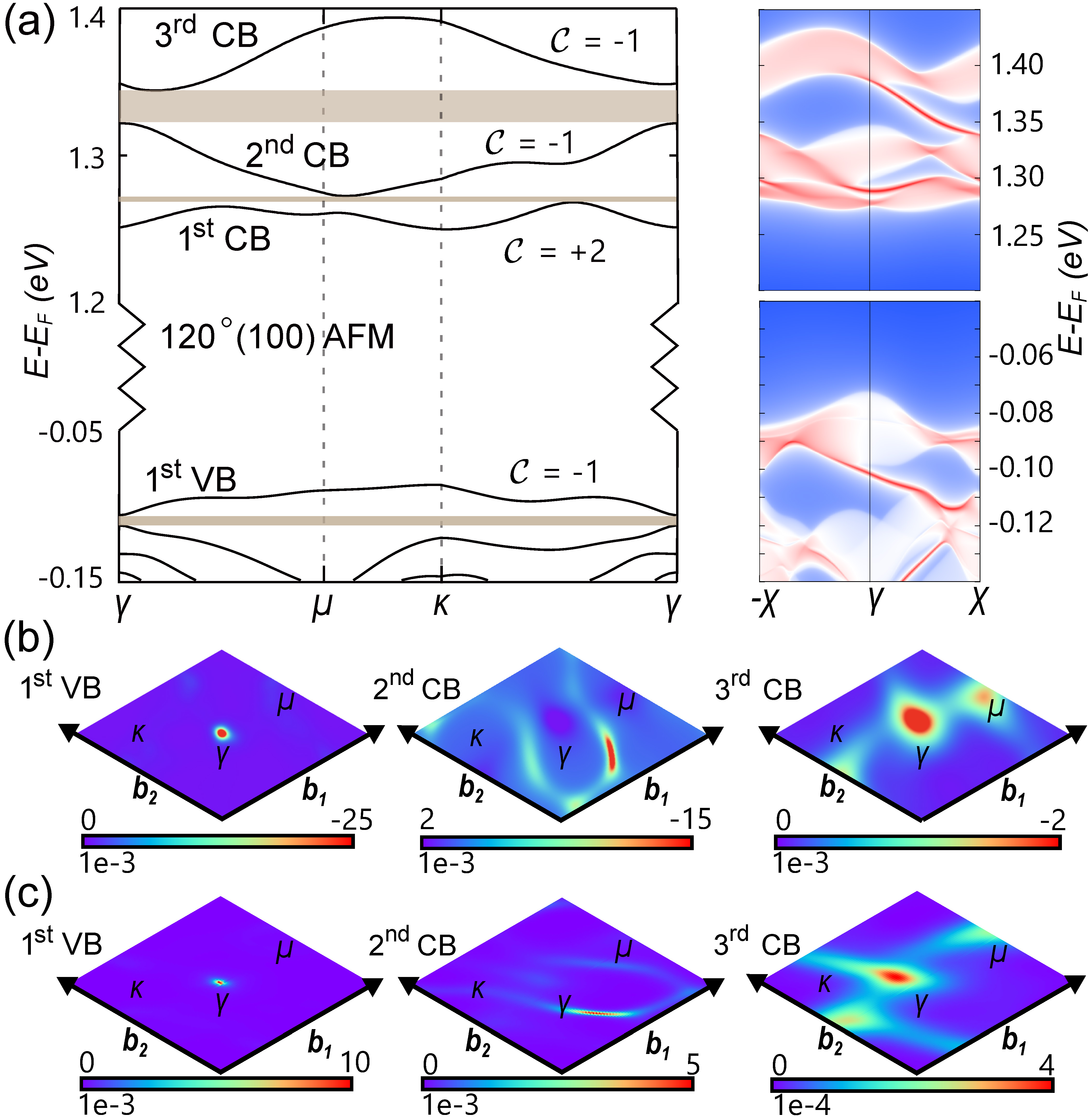}
\end{center}
\caption{Electronic structure and topological properties of monolayer ReAg$_2$Cl$_6$. (a) Band structure and the topological edge states for 120$^{\circ}$ (100) AFM state. Four isolated flat  Chern bands are highlighted. (b,c) The distribution of Berry curvature $\mathcal{B}(\mathbf{k})$ and $\text{Tr}[g(\mathbf{k})]$ in the BZ for three $\mathcal{C}=-1$ Chern bands in (a), respectively. $\mathcal{B}(\mathbf{k})$ remains the same sign throughout the whole BZ for VB and the third CB, while their sign is not always negative for the second CB.}
\label{fig2}
\end{figure}

\emph{Electronic structures and band geometry---}Fig.~\ref{fig2}(a) shows the electronic structure of the $120^{\circ}$ (100) AFM state in ReAg$_2$Cl$_6$. Remarkably, the lowest three conduction bands (CB) and the top valence band (VB) near the Fermi level form \emph{isolated and nearly flat} Chern bands. Their bandwidths and Chern numbers are summarized in Table~\ref{tab2}. The small bandwidths originate from the enlarged unit cell of the magnetic superlattice. Crucially, the nontrivial topology arises from the interplay between $\{a_{1g},e_{g}\}$ orbital band folding, coplanar spin texture, and strong SOC—a mechanism fundamentally distinct from the layer-pseudospin skyrmion lattice in moir\'e MoTe$_2$~\cite{wu2019,devakul2021,yu2020giant,reddy2023,zhang2024polarization}. The Berry curvatures [Fig.~\ref{fig2}(b)] confirm this nontrivial topology, consistent with the chiral edge states observed within the corresponding gaps in the edge local density of states [Fig.~\ref{fig2}(a)].

The interaction energy scale can be estimated as $U\sim e^2/\epsilon a$, where $\epsilon$ is the dielectric constant. Taking $\epsilon=10$, we obtain  $U\sim0.12$~eV. For the isolated flat Chern bands in ReAg$_2$Cl$_6$, the bandwidths are much smaller than $U$, yielding $U/W\gtrsim2$. This dominance of interaction energy establishes favorable conditions for correlated phases. To further assess their suitability for fractionalized states at partial filling, we evaluate two standard band-geometry indicators~\cite{parameswaran2012,roy2014,claassen2015,ozawa2021,mera2021,wang2021,ledwith2023,fujimoto2025}: the Berry curvature fluctuation $\delta\mathcal{B}$ and the average trace condition $\mathbb{T}$, defined as
\begin{eqnarray}
(\delta\mathcal{B})^2 &\equiv& \frac{\Omega_{\text{BZ}}}{4\pi^2}\int_{\text{BZ}} d\mathbf{k} \left(\mathcal{B}(k)-\frac{2\pi\mathcal{C}}{\Omega_{\text{BZ}}}\right)^2,
\\
\mathbb{T} &\equiv& \frac{1}{2\pi}\int_{\text{BZ}} d\mathbf{k}\mathrm{Tr}\left[g(\mathbf{k})\right],
\end{eqnarray}
where $\mathcal{B}(\mathbf{k})\equiv-2\mathrm{Im}(\eta^{xy})$ is the Berry curvature, $g(\mathbf{k})\equiv\mathrm{Re}(\eta^{\mu\nu})$ the Fubini-Study metric, and $\eta^{\mu\nu}(\mathbf{k})\equiv\langle \partial^\mu u_{\mathbf{k}}| \left(1- |u_{\mathbf{k}}\rangle\langle  u_{\mathbf{k}}|\right) |\partial^\nu u_{\mathbf{k}}\rangle$ the quantum geometric tensor. The Chern number is $\mathcal{C}\equiv(1/2\pi)\int d^2\mathbf{k}\mathcal{B}(\mathbf{k})$, $\Omega_{\text{BZ}}$ is the BZ area. The distribution of $\mathcal{B}(\mathbf{k})$ and $\text{Tr}[g(\mathbf{k})]$ are shown in Fig.~\ref{fig2}(b,c), respectively. For VB and the third CB, $\text{Tr}[g(\mathbf{k})]$ are relatively uniform, with standard deviations of $0.92$ and $0.24$, respectively. The calculated values of 
$\delta\mathcal{B}$ and $\mathbb{T}$ are summarized in Table~\ref{tab2}. For comparison, Landau levels with index $\ell$ satisfy $\mathbb{T}=2\ell+1$. Notably, the values obtained for these Chern bands are comparable to those reported in moir\'e materials~\cite{ledwith2020,wang2024,xu2024}, underscoring their suitability for hosting fractionalized topological phases.

\begin{table}[t]
\caption{Bandwidth ($W$), Chern number $\mathcal{C}$, fluctuation of Berry curvature $\delta\mathcal{B}$, and average trace condition $\mathbb{T}$ for isolated Chern bands of 120$^{\circ}$ $(100)$ AFM ground state in ReAg$_2$Cl$_6$.}
\renewcommand{\arraystretch}{1.3}
\begin{center}\label{tab2}
\begin{tabular*}{\columnwidth}
{@{\extracolsep{\fill}}ccccc}
\hline
\hline
Band index & $W$ (meV) & $\mathcal{C}$ & $\delta\mathcal{B}$ & $\mathbb{T}$ \\
\hline
1$^{\rm{st}}$ VB &  $20.8$ & $-1$  & $5.10$ & $2.27$ \\
1$^{\rm{st}}$ CB & $22.7$ & $+2$ & $3.81$ & $2.87$ \\ 
2$^{\rm{nd}}$ CB & $32.1$ & $-1$ & $4.13$ & $4.08$ \\
3$^{\rm{rd}}$ CB & $50.2$ & $-1$ & $1.07$ & $1.66$ \\ 
\hline
\hline
\end{tabular*}
\end{center}
\end{table}

\emph{Correlated states---}To further investigate correlated states in the partially filled Chern bands of the coplanar AFM phase, we construct maximally localized Wannier functions (MLWFs)~\cite{marzari1997,marzari2012,souza2001,brouder2007} and perform many-body calculations based on these Wannier functions. The single-particle Hamiltonian is obtained by projecting the relevant Bloch states near the Fermi level onto MLWFs, which are primarily composed of $\{a_{1g},e_{g}\}$ orbitals of Re. The interacting Hamiltonian is taken as $\mathcal{H}_{\text{int}} = U\sum_{n,i,j,\sigma,\sigma'}^{(i,\sigma)\neq (j,\sigma')}\hat{\rho}^\dagger_{ni\sigma} \hat{\rho}_{nj\sigma'}$, where only onsite interactions between different orbitals and spins are retained, with $c^\dagger_{ni\sigma}$ ($c_{ni\sigma}$) creating (annihilating) an electron of orbital $i$ and spin $\sigma$ on site $n$. The condition $(i,\sigma)\neq (j,\sigma')$ enforces the Pauli exclusion principle, and $U$ denotes the interaction strength. We then carry out exact diagonalization (ED) for the $\mathcal{C}=-1$ Chern bands, namely the VB, the second and third CB, to explore the possible emergence of Abelian fractionalized states. To render many-body calculation tractable, we restrict the variational Hilbert space to the target band and neglect contributions from fully filled lower bands.

\begin{figure}[t]
\begin{center}
\includegraphics[width=\columnwidth,clip=true]{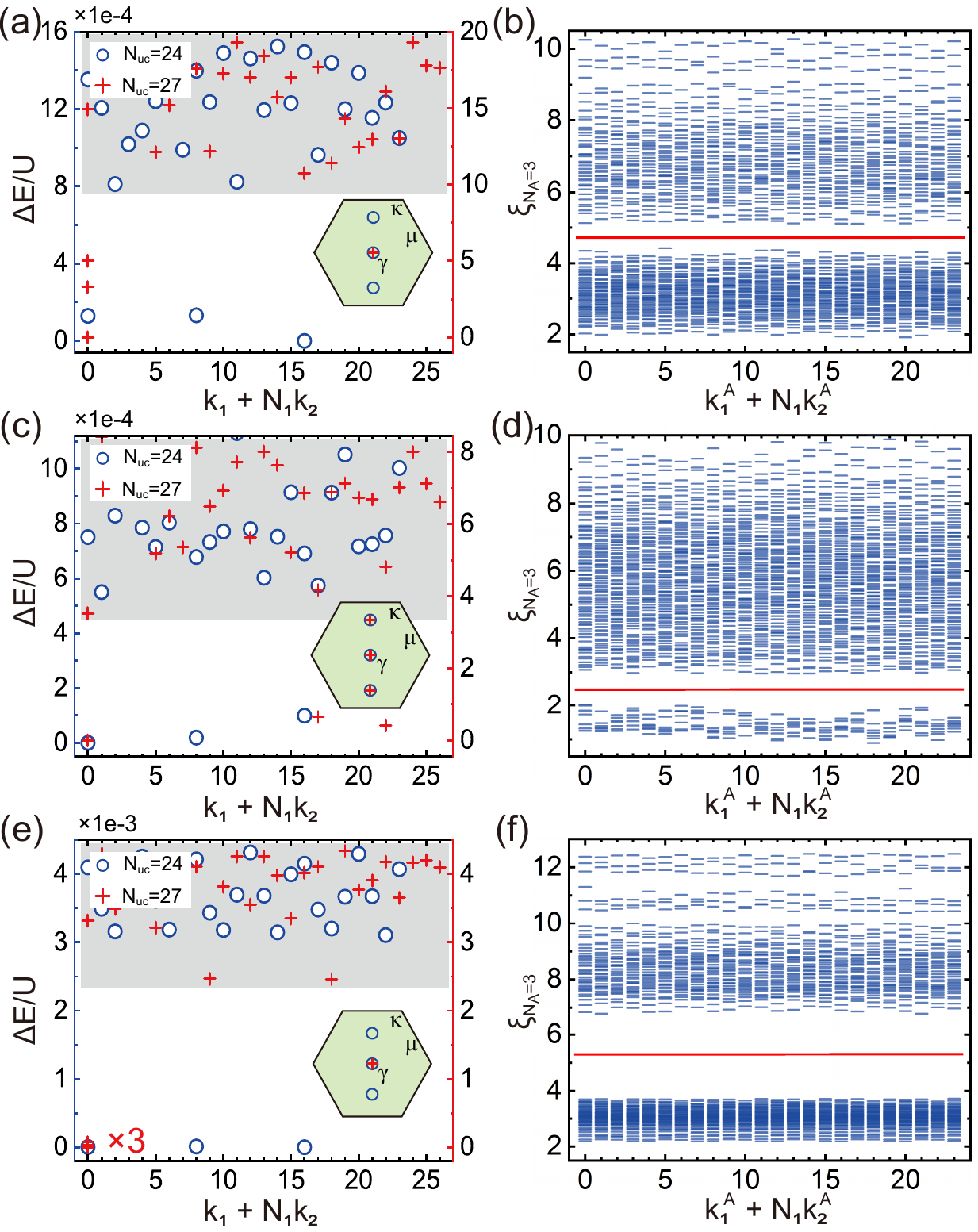}
\end{center}
\caption{Low energy many-body spectra from ED and PES for $1/3$-filled (a,b) 1$^{\rm{st}}$ VB; (c,d) 2$^{\rm{nd}}$ CB; (e,f) 3$^{\rm{rd}}$ CB. ED with $N_{\text{uc}}=24$ and $27$. Insets of ED show the corresponding locations of nearly degenerate ground states of two cluster sizes (marked by blue circles and red crosses) in the BZ. PES with $N_{\text{uc}}=24$ and $N_{A} = 3$ for the three degenerate ground states in (a,c,e). Here we only show the lowest energy per momentum sectors in addition to the degenerate ground state.}
\label{fig3}
\end{figure}

Fig.~\ref{fig3}(a,c,e) show the many-body spectra at filling $\nu=1/3$ for each target band as a function of crystal momentum $\mathbf{k}=k_1\mathbf{T}_1 + k_1\mathbf{T}_2$, which is labeled as $k=k_1 + N_1k_2$. Here $k_{1,2}=0,...,N_{1,2}-1$ for system size $N_{\text{uc}}=N_1\times N_2$ with filled particle number $N_e=\nu N_{\text{uc}}$, $\mathbf{T}_i$ are basis vectors of crystal momentum. We perform calculations for two cluster sizes $N_{\text{uc}}=4\times6$ and $3\times9$. In all three Chern bands, both sizes yield three nearly degenerate ground states that are well separated from excited states (shaded regions in Fig.~\ref{fig3}). The gap persisting across different cluster geometry indicates its stability in the thermodynamic limit. Notably, the momentum sectors of the ground-state manifold for VB [Fig.~\ref{fig3}(a)] and the third CB [Fig.~\ref{fig3}(e)]  are in precise agreement with the generalized Pauli principle, a defining signature of FCI at $\nu=1/3$~\cite{regnault2011}. Moreover, the spectral flow under flux insertion provides extra  evidence of their FCI nature~\cite{supple}. By contrast, the second CB satisfies the Pauli principle only for $N_{\text{uc}}=4\times6$, but not for $N_{\text{uc}}=3\times9$. 

To exclude other competing phases, we further analyze the particle entanglement spectrum (PES) which encodes quasihole excitations~\cite{regnault2011,supple}, by dividing the system into $N_A$ and $N_e-N_A$ particles. We find clear entanglement gaps for degenerate many-body ground states in all three bands with size $N_{\text{uc}}=4\times6$ and $N_{A}=3$ [Fig.~\ref{fig3}(b,d,f)]. For VB and third CB, 
the number of PES levels below the gap exactly matches the counting of quasiparticle excitations resulting from the generalized Pauli principle of Laughlin state. The smaller PES gap in VB compared to the third CB is consistent with its larger fluctuations of 
$\mathrm{Tr}[g(\mathbf{k})]$. By contrast, the number of low-lying levels in the second CB is consistent with the counting rule of CDW. The periodicity of this CDW is determined from the invariant ground-state momenta in ED~\cite{wilhelm2021}, located at $\pm(1/3,1/3)$ [Fig.~\ref{fig3}(c)]. Our ED results demonstrate that partially filled flat Chern bands in ReAg$_2$Cl$_6$ host distinct correlated phases: the VB and third CB stabilize FCI at $\nu=1/3$, while $1/3$-filled second CB supports a commensurate $3\times3\times1$ CDW.

\begin{figure}[t]
\begin{center}
\includegraphics[width=\columnwidth,clip=true]{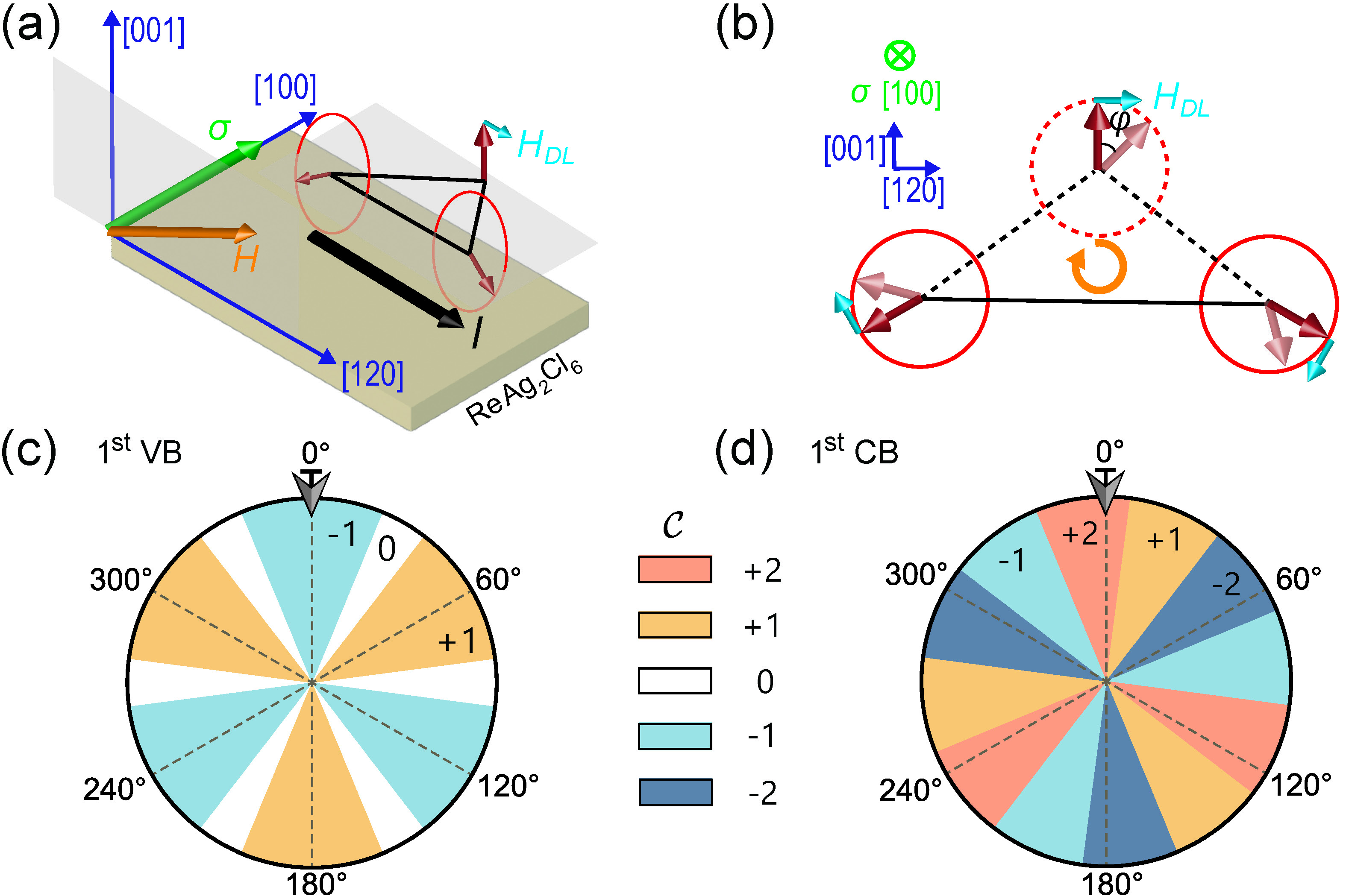}
\end{center}
\caption{SOT switching of topology. (a) Schematic of the experimental setup. A spin current $\boldsymbol{\sigma}$, generated via the spin Hall effect of the applied charge current $\bm{I}$ in the substrate, flows parallel to the monolayer and perpendicular to the magnetic easy plane. The resulting damping-like SOT effective fields $\bm{H}_{\text{DL}}$ act on the sublattice moments within the magnetic easy plane. (b) Illustration of the SOT-driven rotation of sublattice moments in the 120$^{\circ}$ $(100)$ spiral AFM. (c,d) Evolution of the Chern number of VB and the first CB during the spin rotation. The arrows correspond to the spin configurations shown in Fig.~\ref{fig1}(d).}
\label{fig4}
\end{figure}

\emph{Electrically tunable Chern bands}—The band topology is highly sensitive to the spin configuration, whose orientation can be electrically tuned via spin–orbit torque (SOT)~\cite{tsai2020electrical,yoon2023handedness,rimmler2025non,zheng2025all,takeuchi2021chiral,higo2022perpendicular}. As shown in Fig.~\ref{fig4}(a,b), injection of a spin current along $[100]$ induces a damping-like field $\bm{H}_{\text{DL}}$, which rotates the 120$^{\circ}$ coplanar spin texture around the $[100]$ axis. During this rotation, these flat Chern bands remain nearly dispersionless but undergo distinct topological transitions: the VB Chern number evolves from $\mathcal{C}=-1\rightarrow0\rightarrow+1$, while the first CB switches sequentially among $\mathcal{C}=+2,+1,-2,-1$ [see Fig.~\ref{fig4}(c,d)]~\cite{supple}. A 180$^{\circ}$ rotation (time reversal) reverses the Chern number, whereas a 120$^{\circ}$ rotation (sublattice permutation) leaves it invariant. This electrically driven SOT control provides a non-volatile and reversible means to manipulate topological flat bands in 2D coplanar AFMs.

The flat Chern bands in ReAg$_2$Cl$_6$ arise from cooperative effects of coplanar magnetism and topology entirely within the Re $5d$ orbitals. Specifically, the coplanar AFM order quenches the kinetic energy via band folding, while strong SOC converts this magnetic background into nontrivial topology. This mechanism is generic, extending to the ReAg$_2X_6$, ReCu$_2X_6$ and ReAu$_2X_6$ ($X=$ Cl, Br, I) family, which share the same $P$-$3$ lattice symmetry and display similar electronic structures in DFT calculations~\cite{supple}. Further tunability can be achieved by substituting Re with Os or W, thereby introducing an additional $5d$ electron or hole. Monolayer OsAg$_2$Cl$_6$ favors an easy-plane FM ground state, while WAg$_2$Cl$_6$ stabilizes a $120^\circ$ $(001)$ spiral AFM; in both cases, inclusion of SOC drives a QAH phase with FM alignment along [001]~\cite{supple}. Notably, bulk OsAg$_2$Cl$_6$ has already been synthesized~\cite{gromilov2000}, underscoring the experimental feasibility of this material family.

\emph{Discussions---}Realizing these correlated states experimentally requires carrier doping into the Chern bands without compromising the topology. This is achievable via electrostatic gating or chemical substitution. For the spiral AFM state, $1/3$ hole doping into the VB corresponds to a carrier density of order $10^{13}$~cm$^{-2}$~\cite{supple}, which is well within the capability of conventional solid-state and ionic liquid gating~\cite{wu2023electrostatic,guan2023ionic}. Controlled substitution of Re by Os or W offers an effective alternative route for electron or hole doping. DFT calculations show that (Os/W)$_{1/3}$Re$_{2/3}$Ag$_2$Cl$_6$ preserves the (100) AFM ground state and yields an AFM QAH phase, with the band structure nearly intact and the Fermi level shifted only between the first and second CB or VB relative to ReAg$_2$Cl$_6$~\cite{supple,freysoldt2014,wei2004}. 
We thus anticipate that (Os/W)$_x$Re$_{1-x}$Ag$_2$Cl$_6$ with $x\leq1/3$ will retain the same magnetic ground state and the essential features of the electronic structures.

The mechanism of coplanar AFM driving flat Chern bands is not limited to the triangular lattice focused upon here, but can also arise in other geometrically frustrated lattices, such as kagome-lattice systems (e.g., Mn-based alloys). Previous studies~\cite{Messio2011,Chen2014,Feng2015,Nakatsuji2015,Ikhlas2017,Li2017,Zhou2023} have addressed the anomalous Hall effect in coplanar AFMs with zero net magnetization, typically manifesting a metallic state. This stands in stark contrast to our findings of the QAH phase and the subsequent correlated insulating states within flat Chern bands at integer and fractional band fillings, respectively.

In summary, we report a pristine 2D monolayer featuring nearly flat Chern bands stabilized by coplanar spin spirals and strong SOC, which applies to a broad family of Re-based compounds. The fractional fillings of these flat bands are poised to realize FCI and CDW, presenting a new avenue distinct from moir\'e systems. These materials also allow experimental exploration of fractionalized phases in higher Chern bands~\cite{liu2012,wang2012,bar2012,wu2013,sterdy2013,zhang2013}. This family provides a realistic and versatile playground for accessing strongly correlated quantum states in flat topological bands with sizable interactions under ambient conditions.

\begin{acknowledgments}
\emph{Acknowledgments---}We thank J. Gao and Y. Zhang for valuable discussions and sharing data prior to publication. This work is supported by the Natural Science Foundation of China through Grants No.~12350404 and No.~12174066, the Innovation Program for Quantum Science and Technology through Grant No.~2021ZD0302600, the Science and Technology Commission of Shanghai Municipality under Grants No.~23JC1400600, No.~24LZ1400100 and No.~2019SHZDZX01, and is sponsored by the ``Shuguang Program'' supported by the Shanghai Education Development Foundation and Shanghai Municipal Education Commission. K.B. and R.S. contributed equally to this work.
\end{acknowledgments}

\end{document}